\documentstyle[twoside,fleqn,espcrc2,epsfig]{article}

\title{Condensation of vortices and disorder
parameter in 3d Heisenberg model}

\author{A. Di Giacomo\address{Dipartimento di Fisica Universit\`a and INFN
Pisa (Italy)}
        \thanks{Speaker at the Conference.},
        D. Martelli\address{ISAS, Trieste (Italy)},
     G. Paffuti$^a$
}
\begin{document}

\begin{abstract}
The 3d Heisenberg model is studied from a dual point of view in terms of 2d
solitons (vortices). It is shown that the disordered phase corresponds to condensation of
vortices in the vacuum, and the critical indices are computed from the
corresponding disorder parameter.
\end{abstract}

\maketitle

\section{Introduction}
The Heisenberg ferromagnet is defined by the partition function
\begin{equation}
Z[\beta] = \int \prod [d \Omega(x) \exp(-S)\label{eq:1}\end{equation}
where
\begin{equation}
S = \frac{1}{2}\beta \sum_{\mu,x}\left[ \Delta_\mu \vec n(x)\right]^2
\qquad \vec n^2(x) = 1\label{eq:2}\end{equation}
and $d\Omega(x)$ is the element of solid angle for the orientation of $\vec n$
in colour space.

The model presents a 2nd order phase transition at $\beta_c\simeq 0.7$\cite{1}.
For
$\beta>\beta_c$ there is an ordered phase, with order parameter the
magnetization $\langle\vec n\rangle\neq 0$; for $\beta<\beta_c$,
$\langle\vec n\rangle = 0$ (disordered phase). We shall describe the system
from a dual point of view, and show that in the disordered phase vortices
condense.
The model will be viewed as a $2+1$ dimensional euclidean field theory.
Vortices will be labelled by the integer valued topological charge of the 2
dimensional spacial configurations, which is a conserved quantity, and defines
a $U(1)$ symmetry of the system. A disorder parameter $\langle\mu\rangle$ will
be constructed, which detects spontaneous breaking of this $U(1)$ symmetry, or
condensation of vortices.

The situation is
analogous to the condensation of monopoles in the confining phase of gauge
theories\cite{2,3} or to the condensation of abelian vortices in
$He_4$\cite{4}. To check our construction we shall extract from the numerical
determination of
$\langle\mu\rangle$ at the phase transition the known critical indices.
\section{The Heisenberg model as a fiber bundle.}
Usually the colour frame to which the direction of $\vec n$ is referred is a
fixed frame, independent of $x$
\[
\vec\xi_i^0\,(i=1,2,3)\quad
\vec\xi_i^0\wedge\vec\xi_j^0 = \vec\xi_ki^0\quad
\vec\xi_i^0\cdot\vec\xi_j^0 = \delta_{ij}\]
A body fixed frame can be defined\cite{6} by three unit vectors
$\vec\xi_i(x)\,(i=1,2,3)$
\[
\vec\xi_i(x)\wedge\vec\xi_j(x) = \vec\xi_ki(x)\quad
\vec\xi_i(x)\cdot\vec\xi_j(x) = \delta_{ij}\]
and $\vec\xi_3(x) \equiv \vec n(x)$. The frame is defined up to an arbitrary
rotation around $\vec n(x) = \vec\xi_3(x)$.

Since $\vec \xi_i^2 = 1$
\[ \partial_\mu \vec\xi_i(x) = \vec\omega_\mu\wedge \vec\xi_i(x)\]
or
\begin{eqnarray}
D_\mu\vec\xi_i(x) &\equiv& (\partial_\mu - \vec\omega_\mu\wedge)\vec\xi_i = 0
\label{eq:3}\\
D_\mu &=& \partial_\mu - i\,\vec\omega_\mu\cdot\vec T\nonumber
\end{eqnarray}
$(T^a)_{ij} = i \varepsilon_{iaj}$ are the generators of the $O(3)$ symmetry
group.
Eq.(\ref{eq:3}) is nothing but the definition of parallel transport. From
eq.(\ref{eq:3}) it follows $\left[D_\mu,D_\nu\right]\vec\xi_i = 0$ or, by
completeness of
$\vec\xi_i$
\begin{equation}
[D_\mu,D_\nu] = \vec T\cdot\vec F_{\mu\nu}(\omega) = 0
\label{eq:4}\end{equation}
\[ \vec F_{\mu\nu}(\omega) =
\partial_\mu\vec\omega_\nu - \partial_\nu\vec\omega_\mu +
\vec\omega_\mu\wedge\vec\omega_\nu\]
$\vec\omega_\mu$ is a pure gauge, apart from singularities. The general
solution of eq.(\ref{eq:3}) is then, for $\vec n(x) = \vec\xi_3(x)$
\[\vec n(x) = P\,\exp
\left(i\,\int_{\infty,C}^x\vec T \vec \omega_\mu(x') d x'_\mu\right) \vec n_0
\]
where $\vec n_0$ is the value of $\vec n(x)$ at infinity, and the dependence on
the path $C$ is trivial, because $F_{\mu\nu}$ is a pure gauge, eq.(\ref{eq:4}).
This is true apart from singularities. We will show that such singularities
exist.

The current
$ J_\mu = \frac{1}{8\pi}\varepsilon_{\mu\alpha\beta}\vec n
\cdot\left(\partial_\alpha\vec n\wedge\partial_\beta\vec n\right)$
is identically conserved
\begin{equation}
\partial^\mu J_\mu = 0 \label{eq:4a}\end{equation}
If we look at the theory as the euclidean version of a field theory, with
euclidean time on the 3 axis, the corresponding conserved quantity is
\[ Q = \frac{1}{4\pi}
\int d^2 x\vec n\cdot\left(\partial_1\vec n\wedge\partial_2\vec n\right)\]
which is nothing but the topological charge of the
2 dimensional configurations of the theory.
$Q$ can assume positive and negative integer values. By use of
eq.(\ref{eq:3}) and eq.(\ref{eq:4}) it is easy to show that
\begin{equation}
Q =
\frac{1}{4\pi}
\int d^2 x\vec n\cdot\left(\vec\omega_1\wedge\vec\omega_2\right)=
-
\oint_C(\vec\omega_i\cdot\vec n)d x^i \label{eq:5}\end{equation}
where the path $C$ is the contour of the region
in the 2 dimensional space ($x_3$ = const.) where eq.(\ref{eq:4}) holds.
Since $Q=\pm n$
Eq.(\ref{eq:5}) shows that $\vec \omega_\mu$ is not always a pure gauge.

This can be explicitely checked on a configuration corresponding to a static 2
dimensional instanton propagating in time (vortex).

The conserved current $J_\mu$ identifies a $U(1)$ symmetry. We will show that
this symmetry is Wigner in the ordered phase $\beta>\beta_c$, and is
spontaneously broken in the disordered phase.
\section{Disorder parameter}
Let $R_q(\vec x,\vec y)$ be a $\vec x$ dependent singular rotation creating a
vortex of charge $q$ at the site $\vec y$ in a 2 dimensional configuration.
\[ R_q(\vec x,\vec y) \vec n(\vec x,t) \Rightarrow
\vec n_q(\vec x,t) \]
The creation operator of a vortex $\mu_q$ at site $\vec y$, time $t$, will be
defined as
\begin{eqnarray}&&\hskip-25pt
\mu_q(\vec y,t)\! =
\exp\Bigl\{\!
-\beta\!\sum_x\!\bigl[(R_q^{-1}(\vec x,\vec y)\vec n(\vec x,t+1) - \vec n(\vec
x,t))^2
\nonumber\\
&&-(\vec n(\vec x,t+1) - \vec n(\vec x,t))^2\bigr]\Bigr\}
\label{eq:7}\end{eqnarray}
We measure the correlator
\[
{\cal D}(x_0)=
\langle \mu_{-q}(\vec 0,x^0)\mu_q(\vec 0,0)\rangle\]
By cluster property
\begin{equation} {\cal D}(x_0)
\mathop\simeq_{|x_0|\to\infty}
Ae^{-M|x_0|} +\langle\mu_q\rangle^2\label{eq:8}\end{equation}
$\langle\mu_q\rangle\neq0$ signals spontaneous breaking of the $U(1)$ symmetry
(\ref{eq:4a}), or condensation of vortices.

By use of the definition
 (\ref{eq:7}) it is easy to see that
\[ {\cal D}(x_0) = \frac{ Z[S+\Delta S]}{Z[S]}\]
where
$S+\Delta S$ is obtained from $S$, eq.(\ref{eq:2}) by the replacement
\begin{eqnarray*}
\left[\Delta_0\vec n(\vec x,0)\right]^2 \hskip-5pt&\to&\hskip-5pt
\left[R_q^{-1}(\vec x,0)\vec n(\vec x,1) - \vec n(\vec x,0)\right]^2
\\
\left[\Delta_0\vec n(\vec x,x_0)\right]^2 \hskip-5pt&\to&\hskip-5pt
\left[R_q^{-1}(\vec x,0)\vec n(\vec x,x_0+1) - \vec n(\vec x,x_0)\right]^2
\end{eqnarray*}
and that this really amounts to have a vortex propagating from $(\vec0,0)$ to
$(\vec 0,x_0)$\cite{2}.

Instead of ${\cal D}(x_0)$ itself it proves convenient\cite{2,3} to study the
quantity $\rho(x_0) = \frac{1}{2} d\ln{\cal D}(x_0)/d\beta$. As
$|x_0|\to\infty$ from eq.(\ref{eq:8}) we have for
$\rho \equiv \rho(x_0 = \infty)$, $\rho\simeq d \ln\langle\mu\rangle/d\beta$. Since
$ \langle\mu\rangle_{\beta=0} = 1$, $ \langle\mu\rangle = \exp\left[
\int_0^\beta \rho(\beta') d \beta'\right]$. $\rho$ is easier to measure and
contains all the information on the transition. The behaviour of $\rho$ is
shown in fig.1.

For $\beta < \beta_c$, $\rho\to$~finite limit consistent with zero, or
$\langle\mu\rangle\neq 0$,
which means condensation of vortices.

For $\beta > \beta_c$ $\rho$ can be evaluated in
perturbation theory and behaves as
\begin{equation}
\rho = - c_1 L + c_2 \qquad c_1>0 \label{eq:9}\end{equation}
$L$ is the lattice size. Eq.(\ref{eq:9}) implies that $\langle\mu\rangle = 0$
for $\beta > \beta_c$ in the thermodynamical limit $V\to\infty$.
Around $\beta_c$ a finite size scaling analysis can be performed
\[\langle \mu\rangle = f\left(\frac{a}{\xi},\frac{L}{\xi}\right) \sim
f\left(0,\frac{L}{\xi}\right)\]
\par\noindent
\begin{minipage}{0.5\textwidth}
\epsfig{figure=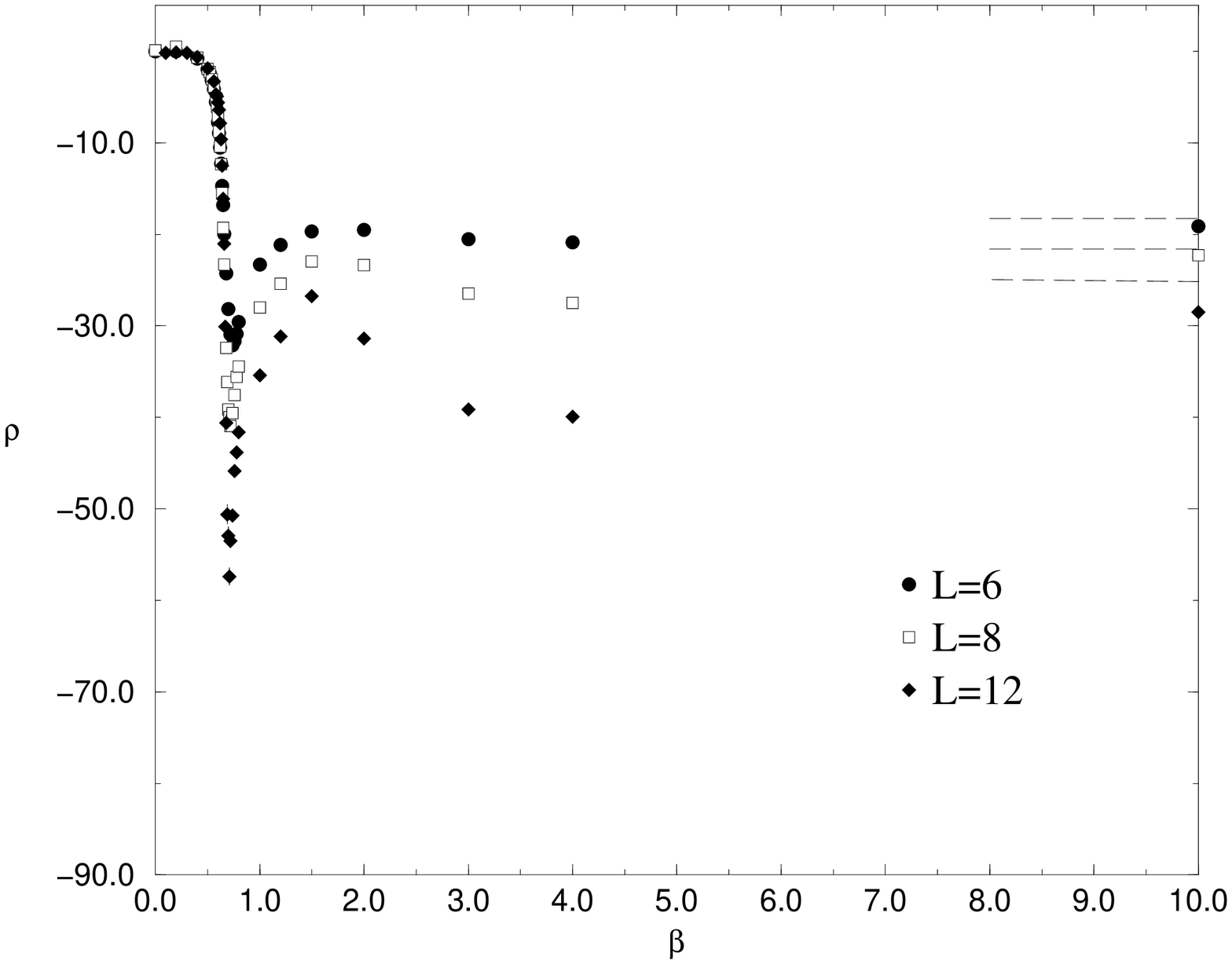,  width = 0.9\textwidth, angle=270}
\vskip0.1in
{\centerline{\bf Fig.1}}
\end{minipage}
\vskip0.1in
\par\noindent
and since $1/\xi\simeq (\beta_c-\beta)^\nu$, we get the scaling law\cite{2}
\[ \frac{\rho}{L^{1/\nu}} = f(L^{1/\nu}(\beta_c-\beta))\]
The scaling law is verified, fig.2, and allows to extract $\beta_c$ and $\nu$
\begin{eqnarray*}
\beta_c &=& 0.695\pm0.003\; [0.6928]\\
 \nu &=& 0.70 \pm 0.02\; [0.698]\end{eqnarray*}
They agree with the values determined from $\langle\vec n\rangle$, which are
indicated in parentheses\cite{1,6a}.
\par\noindent
\begin{minipage}{0.5\textwidth}
\hskip-10pt\epsfig{figure=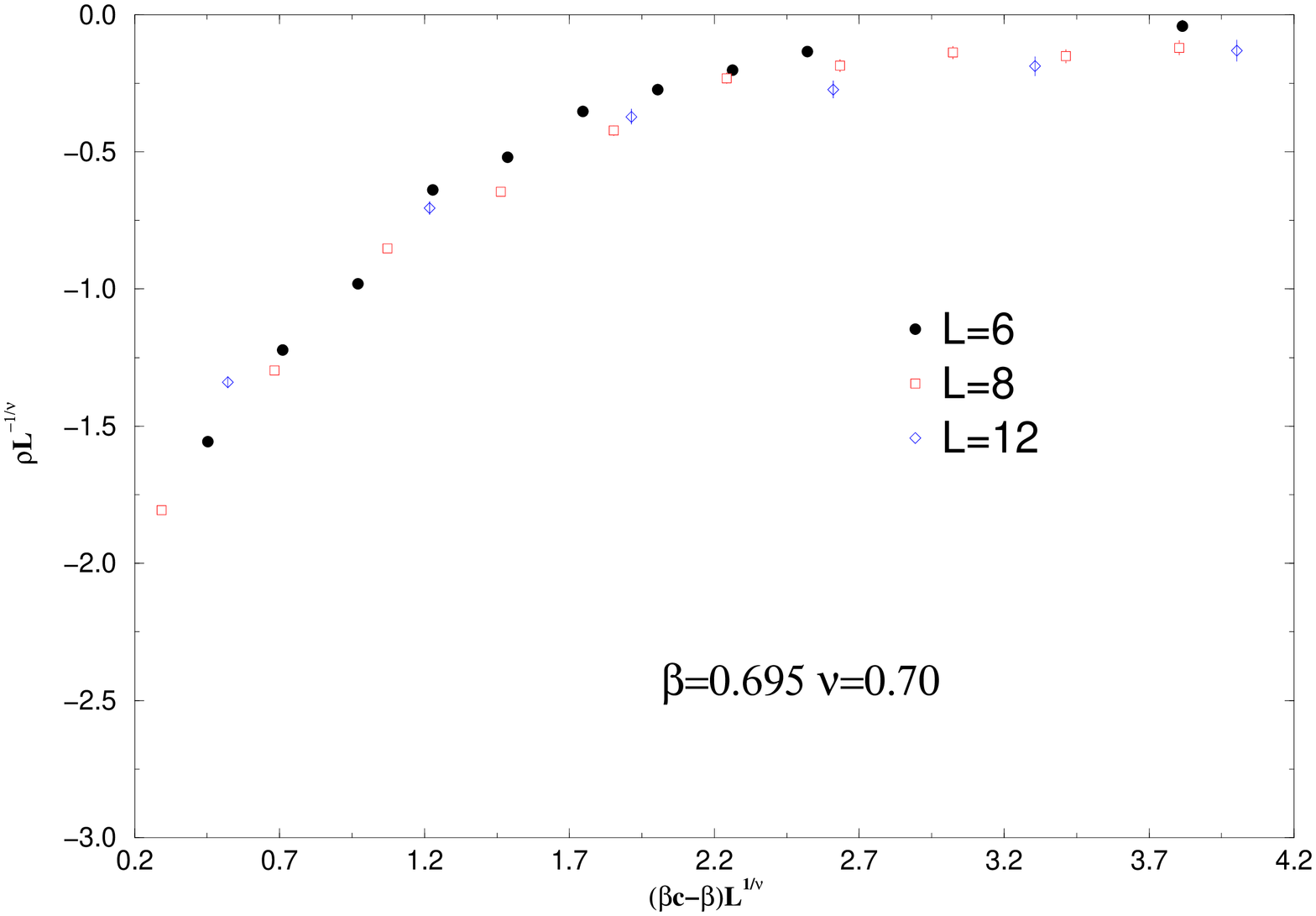,  width = 0.75\textwidth, angle=270}
\vskip0.1in
{\centerline{\bf Fig.2}}
\end{minipage}
\section*{Conclusions}
The phase transition to disorder in 3d Heisenberg model is produced by
condensation of topological solitons. A disorder parameter can be defined and
out of it the critical indices can be determined.
\section*{Acknowledgements}
This work has partially supported by MURST. A. Di Giacomo is grateful to EC,
TMR project, ERBFMX-CT97-0122 for financing the partecipation to the conference.


\begin{thebibliography}{9}
\bibitem{1} E. Brezin, J. Zinn-Justin, {\em Nucl. Phys.} {\bf
B257}, 867, (1985).
\bibitem{2} A. Di Giacomo, G. Paffuti, {\em Phys. Rev.}
{\bf D56}, 6816, (1997).
\bibitem{3}L.Del Debbio, A.Di Giacomo, G.Paffuti and P.Pieri,
{\it Phys. Lett.}{\bf B 355} (1995) 255.
\bibitem{4} G. Di Cecio, A. Di Giacomo, G. Paffuti, M. Trigiante,
{\em Nucl. Phys.} {\bf B489}, 739, (1997).
\bibitem{6}A. Di Giacomo, M. Mathur,
{\em Phys. Lett.} {\bf B400}, 129, (1997).
\bibitem{6a} P. Peczak, A.L. Ferrenberg, D.P. Landau,
{\em Phys. Rev.} {\bf 43}, 6087, (1991).
\end{thebibliography}
\end{document}